\documentclass[12pt,preprint]{aastex} 

\shorttitle{A Photometric Diagnostic to Aid in the Identification of 
Transiting Extra-Solar Planets} 
\shortauthors{Tingley \& Sackett} 
 
\begin{document} 
 
\title{A Photometric Diagnostic to Aid in the Identification of 
Transiting Extra-Solar Planets} 
 
\author{Brandon Tingley and Penny D. Sackett} 
\affil{Research School of Astronomy and Astrophysics, and \\
Planetary Science Institute, \\
The Australian National University, Mount Stromlo Observatory, Cotter Road,\\ 
Weston, Canberra ACT 2611, Australia}
 
\begin{abstract} 
One of the obstacles in the search for exoplanets via transits is the large
number of candidates that must be followed up, few of which ultimately prove
to be exoplanets. Any method that could make this process more efficient by
somehow identifying the best candidates and eliminating the worst would
therefore be very useful. \citet{seager} demonstrated that it was possible to
discern between blends and exoplanets using only the photometric
characteristics of the transits. However, these techniques are critically
dependent on the shape of the transit, characterization of which requires very
high precision photometry of a sort that is atypical for candidates identified
from transit searches. We present a method relying only on transit duration,
depth, and period, which require much less precise photometry to determine
accurately. The numerical tool we derive, the exoplanet diagnostic $\eta$, is
intended to identify the subset of candidates from a transit search that is
most likely to contain exoplanets, and thus most worthy of subsequent
follow-up studies. The effectiveness of the diagnostic is demonstrated with
its success in separating modeled exoplanetary transits and interlopers, and by
applying it to actual OGLE transit candidates.
\end{abstract} 
 
 
\section{Introduction} 
 
One of the fundamental problems in the search for exoplanets via transits is
the relatively small fraction of systems exhibiting periodic transits that are
found to be caused by exoplanets after follow-up observations reveal the
nature of the orbiting body. The process of verification consumes a great deal
of telescope time on top-class instruments. This process would be made more
efficient by the development of a method that could identify the best and
worst candidates from the collection of blends, binaries and exoplanets that
comprise a typical candidate set from transit searches. \citet{seager} touched
on this possibility while exploring the feasibility of estimating stellar
parameters from the photometric properties of an observed transit. Using a
fully analytical derivation, they showed that four parameters (depth, period,
full transit duration, and duration minus ingress and egress times) could be
used to calculate physical parameters such as mean stellar density and radius
ratio for the system, assuming that the transiting body is an exoplanet and
that the orbit is circular. Non-planetary transiting bodies will produce
somewhat aphysical parameters, particularly in the radius of the transiting
body. A typical blend, for example, would produce a radius for the transiting
body that would be 20\%-50\% larger than that expected for a close-in giant
planet.
 
Unfortunately, this technique requires extremely precise photometry. Seager \&
Mall\'{e}n-Ornelas quote a need for 5 millimag precision with 5 minute
sampling for two transits (provided these two transits define the true period)
for short-period giant planets.  Current transit search campaigns do not
produce data of this quality, which limits the applicability of this
technique.
 
However, as we show, it is still possible to use the photometric properties of
transits derived from less precise data to identify the best exoplanet
candidates from a large sample. By using a set of reasonable approximations,
we derive a numerical tool (called the exoplanet diagnostic) that indicates
how ``planet-like'' a particular event is using only the transit period,
duration and depth. This diagnostic makes it possible to exclude many of the
candidates from transit searches without the need for follow-up observations,
including many of those caused by blends.
 
In Section 2, we derive the exoplanet diagnostic and discuss the impact of 
orbital eccentricity. In Section 3, the results of our analysis of existing 
transit searches and of the effectiveness of the diagnostic in distinguishing 
modeled blends is presented 
 
\section{Derivation of the Exoplanet Diagnostic $\eta$} 
 
The duration, $D$, of a transit depends on many parameters: the radii and
masses of the two transiting objects, semi-major axis, orbital inclination,
eccentricity, viewing orientation and period. A completely general equation
describing the duration of the transit can be derived from these variables.
We begin with the derivation of transit duration given by \citet{sackett},
 
\begin{equation} 
D = \frac{\Delta\phi}{\omega_{\rm t}} = 
\frac{\Delta\phi \, r_{\rm t}}{v_{\rm t}} ~~~, 
\end{equation} 
 
\noindent 
in which $\Delta\phi$ is the eccentric angle between the first and last
contacts of the transit, $\omega_{\rm t}$ is the angular velocity of the
planet at the time of the transit, $r_{\rm t}$ is the separation between the
planet and the parent star at the time of transit, and $v_{\rm t}$ is the
orbital velocity at this time. Assuming that $r_{\rm t}$ does not change
appreciably during transit, and that $r_{\rm t} \gg R_\star$ (see Fig.~1b),
$\Delta\phi$ can be written as
 
\begin{equation} 
\Delta\phi = \arcsin\left(2\sqrt{\frac{(R_1+R_2)^2}{r_{\rm t}^2}-\cos^2 i}\right) 
\approx 2\sqrt{\frac{(R_1+R_2)^2}{r_{\rm t}^2}-\cos^2 i} ~~~,  
\end{equation} 
 
\noindent 
where $R_1$ and $R_2$ are the radii of the transiting bodies and $i$ is the
orbital inclination.  Invoking conservation of angular moment $L$,
 
\begin{equation} 
L = \mu \sqrt{GM_{\rm tot} \, a(1-e^2)} = \mu \, v_{\rm t} \, r_{\rm t} ~~~,  
\end{equation} 
 
\noindent 
where $M_{\rm tot}$ is the total mass of the system, $\mu \equiv (m_1
m_2)/M_{\rm tot}$ is the reduced mass, $a$ is the semi-major axis of the
orbit, and $e$ is the orbit eccentricity.  Putting these pieces together
yields:
 
\begin{equation} 
D = \frac{2(R_1+R_2) \, 
r_{\rm t}}{\sqrt{GM_{\rm tot} \, a(1-e^2)}} 
\sqrt{1-\frac{r_{\rm t}^2 \cos^2i}{(R_1+R_2)^2}} \equiv 
\frac{2(R_1+R_2) \, r_{\rm t}}{\sqrt{GM_{\rm tot} \, a(1-e^2)}} 
Z ~~~,  
\end{equation} 
 
\noindent 
where we have absorbed the geometrical effects of the projected inclination
into $Z$.  The quantity $Z$ at most unity (in the case of an $i = 90^\circ$
central transit) and increasingly less than unity for increasing degrees of
misalignment (see Fig.~2). In order to eliminate $a$ and $r_{\rm t}$, we invoke Kepler's
third law,
 
\begin{equation} 
a = \left(\frac{GM_{\rm tot} \, \tau^2}{4\pi^2}\right)^{\frac{1}{3}} ~~~,  
\end{equation} 
 
\noindent 
where $\tau$ is the period of the orbit. Then, recalling the equation for an
ellipse
 
\begin{equation} 
r_{\rm t} = \frac{a(1-e^2)}{1+e\cos\phi_{\rm t}} ~~~,  
\end{equation} 
 
\noindent 
where $\phi_{\rm t}$ is the eccentric angle at transit center of the system,
relative to the semi-major axis of the orbit ($\phi_{\rm t}$ depends only on
viewing geometry; see Fig.~1a), we can write the final equation for the
transit duration as
 
\begin{equation} 
D = 2Z \, (R_1+R_2) \frac{\sqrt{1-e^2}}{1+e\cos\phi_{\rm t}} \left(\frac{\tau}{2\pi 
GM_{\rm tot}}\right)^{\frac{1}{3}} ~~~.  
\end{equation} 

\noindent 
This equation is general; in the case of exoplanets, $M_{\rm tot} \approx
M_\star$, the mass of the parent star.
 
\subsection{Circular Exoplanetary Orbits} 
 
As there are a large number of unknowns, we first examine the simplest case of
circular ($e=0$) orbits.  This is a relevant case since transit searches are
most sensitive to "hot Jupiters" (that is, Jovian-mass planets with $a \le
0.1$ AU), all of which discovered thus far have very low eccentricities.  For
$e = 0$ and $M_{\rm tot} = M_\star$, Eq.~7 reduces to
 
\begin{equation} 
D = 2Z \, (R_{\rm p}+R_\star) \left(\frac{\tau}{2\pi GM_\star}\right)^{\frac{1}{3}} 
\end{equation} 
 
\noindent 
where $R_{\rm p}$ is the radius of the exoplanet and $R_\star$ is the radius
of the parent star. This contains four unknowns that cannot be observed
directly: $Z$, $R_{\rm p}$, $R_\star$ and $M_\star$.  However, as we show
below, it is possible to replace all but one of these with observable
quantities.
 
Through extensive modeling of planetary transits using a standard ZAMS and
limb darkening from from \citet{claret} for many different exoplanet-star size
ratios and projected inclinations, we find that the fractional depth $d$ of a
central exoplanetary transit in the $I$ passband can be approximated as
 
\begin{equation} 
d \approx 1.3 \left(\frac{R_{\rm p}}{R_\star}\right)^2 ~~~. 
\end{equation} 
 
\noindent
Off-center transits will have only slightly lower constants of proportionality
unless they are grazing, which is an unlikely, but still possible, scenario.
  
The mass-radius relationship for the lower main sequence \citep{cox} 
proposes: 
 
\begin{equation} 
M_\star \approx M_\odot \left(\frac{R\star}{R_\odot}\right)^{1.25} ~~~. 
\end{equation} 
 
\noindent
With these three equations, we create a diagnostic $\eta$, namely the ratio of
the theoretical duration $D$ (derived above) and the observed duration,
$D_{\rm obs}$ (obtained from photometric observations).  There are still
unknowns: the factor $Z$, known to be constrained to lie between 0 and 1, and
one of $R_\star$ or $R_{\rm p}$. The latter duality leads to two different
forms of the exoplanet diagnostic, $\eta_{\rm p}$ and $\eta_{\star}$, each of
which is useful in different circumstances.
 
Radii of stars that are thought to have a reasonable chance of harboring an
exoplanet, namely main sequence spectral types F through late K, can vary by a
factor of three or so, while the typical planetary radius that can produce a
detectable transit from ground varies by about a factor of two.  Therefore,
ground-based transit searches would be better served by a diagnostic
($\eta_{\rm p}$) that takes the unknown $R_{\rm p}$ to be constant, while
satellite missions, which are sensitive to a much wider range of $R_{\rm p}$,
would be better served using a diagnostic ($\eta_\star$) that assumed
$R_\star$ was constant. We stress, however, that for any primaries for which
the radius can be estimated (from spectral classification, for example),
$\eta_\star$ should be used.  These two exoplanet diagnostics, easily
derivable from the above equations, are:
 
\begin{equation} 
\eta_{\rm p} \equiv \frac{D_{\rm obs}}{D} 
%%\[ \;\;\;\;\;\; = \frac{D_{\rm obs}}{2Z\left(1+\sqrt{\frac{1.3}{d}}\right)}\;\;\;\times \] 
%%\;\;\;\;\;\;\;\;\;\;\;\; \left(\frac{2\pi GM\odot}{\tau}\right)^{\frac{1}{3}} 
%%R_{\rm p}^{\frac{-7}{12}}R_\odot^{\frac{-5}{12}}\left(\frac{1.3}{d}\right)^{\frac{5}{24}} 
= \frac{D_{\rm obs}}{2Z \left(1+\sqrt{1.3/d}\right)} 
\left(\frac{2\pi GM\odot}{\tau}\right)^{\frac{1}{3}} 
R_{\rm p}^{-\frac{7}{12}}R_\odot^{-\frac{5}{12}}\left(\frac{1.3}{d}\right)^{\frac{5}{24}} 
\end{equation} 
 
\noindent 
and 
 
\begin{equation} 
\eta_\star \equiv \frac{D_{\rm obs}}{D} 
= \frac{D_{\rm obs}}{2Z \left(1+\sqrt{d/1.3}\right)} 
\left(\frac{2\pi GM\odot}{\tau}\right)^{\frac{1}{3}} 
R_\star^{-\frac{7}{12}}R_\odot^{-\frac{5}{12}} ~~~. 
\end{equation} 
 
\noindent
Note that $\eta_{\rm p}$ is independent of $R_\star$ and $\eta_\star$ is
independent of $R_{\rm p}$ . The one remaining unknown left in these
equations, $Z$, will not critically affect the viability of these diagnostics.
In what follows, we take $Z = 1$ and describe the consequences of this
assumption.

Looking at Eqs. 4 and 7, one would expect the typical binary transit to last
longer than the typical exoplanet transit, as the radius of a transiting star
would far exceed that of a transiting planet, unless it was an extreme
grazing eclipse. The $R_{\rm p}+R_\star$ for an exoplanetary transit becomes
$R_{\star,1}+R_{\star,2}$ for a stellar eclipsing binary -- a large increase
in this term.

A more in-depth examination of these diagnostics gives greater insight into
the situation. $\eta_{\rm p}$ scales as
 
\begin{equation} 
\eta_{\rm p} \propto D_{\rm obs}d^{\frac{7}{24}} 
R_{\rm p}^{-\frac{7}{12}}\tau^{-\frac{1}{3}} ~~~. 
\end{equation} 

\noindent
while $\eta_\star$ scales as

\begin{equation} 
\eta_{\star} \propto D_{\rm obs}\left(1-\sqrt(\frac{d}{1.3})\right) 
R_{\star}^{-\frac{7}{12}}\tau^{-\frac{1}{3}} ~~~. 
\end{equation} 

\noindent 
Decreasing the inclination of the orbit not only reduces the measured
duration, $D_{\rm obs}$, but also reduces the transit depth, $d$. Both factors
act to reduce $\eta_{\rm p}$, while they counteract each other for
$\eta_\star$. The latter has a much stronger dependence on $D_{\rm obs}$ than
$d$ for the shallow events that are of interest, so decreasing $Z$ also
decreases $\eta_\star$. Following this line of reasoning, the diagnostics will
therefore be a maximum for central transits (with $Z=1$) and less for more
grazing transits. Thus, true exoplanetary events would have a diagnostic less
than or equal to one, while stellar events (blends or grazing binaries) would
often, but not always, have diagnostics greater than one. 

Some grazing stellar binaries and blended eclipsing binaries will have similar
diagnostics to exoplanets, which will lead to some stellar contamination in
the reduced set of candidates even after the diagnostic has been applied.
Some grazing binaries can be identified via a clearly V-shaped transit
\citep{udalski1}, the presence of ellipsoidal variations \citep{drake,sirko},
or by virtue of the existence of a detectable secondary transit that has a
different depth than that of the primary transit.  Also, binary systems in
which both stars are nearly identical will have a photometric period that is
one-half the actual period, as it will have two identical transits per
orbit. Because of this, their diagnostics are increased by a factor of
$2^{\frac{1}{3}}$, helping to push them out of the regime populated by
exoplanetary events. Additionally, a large percentage of systems involving
transits of giant stars could in theory be identified as non-planetary, as
their large radii lead to very long transit durations.
 
The exoplanet diagnostic should also be useful in separating most blends from
exoplanetary transits. Blended eclipsing binaries would have a high
$R_{\star,1}+R_{\star,2}$ compared to exoplanets and the dilution of their
transits to a shallower depth by a third component means that their transits
would generally be more central than a grazing binary for a given transit
depth. They would thus have a longer duration (and a higher exoplanet
diagnostic) relative to other events with a similar transit depth, whether
they be grazing binaries or exoplanets. This is potentially very useful for
missions such as Kepler, which will likely identify large numbers of objects
that exhibit periodic transits of a variety of depths -- many of which will be
blends. The shallowest events are both the most interesting and the most
difficult to confirm via follow-up observations; any technique that can
eliminate some of these without follow-up will save significant astronomical
resources.
 
\subsection{Elliptical Exoplanetary Orbits} 
 
A good fraction of the exoplanets that have been discovered to date have
nearly circular orbits. In fact, the average eccentricity for all known
exoplanets with periods less than about 60 days is approximately 0.1. However,
the majority has periods longer than 60 days, where the average eccentricity
rises sharply to 0.3 \citep{halbwachs}. We are thus led to consider the effect
of eccentricity on the diagnostic. As can be seen from Eq.~7, the diagnostic
depends on the eccentricity in the following fashion:
 
\begin{equation} 
\eta \propto \frac{D_{\rm obs}}{D_{\rm circ}} = 
\frac{\sqrt{1-e^2}}{1+e\cos\phi_{\rm t}} 
\end{equation} 
 
\noindent 
The dependence on the viewing angle $\phi_{\rm t}$ 
creates a distribution in the possible diagnostics for a given period and 
eccentricity given by: 
 
\begin{equation} 
S_1(\eta) = \frac{d\phi_{\rm t}}{d\eta} = 
\frac{d}{d\eta}\left(\arccos\left(\frac{\sqrt{1-e^2}-\eta}{e\eta}\right)\right) 
= \frac{(1-e^2)^\frac{1}{4}}{\eta}\left(2\eta - (1 + \eta^2)\sqrt{1-e^2}\right)^{-\frac{1}{2}} ~. 
\end{equation} 
 
\noindent 
However, the probability that a transit will be observed to occur is inversely
proportional to $r_{\rm t}$ \citep{sackett}.  This means that transits are
more likely to occur when $r_{\rm t}$ is smaller, the orbital velocities are
correspondingly higher, and thus transit durations are correspondingly
shorter.  This selection effect skews the spread in observed $\eta$ toward
lower diagnostics (see Fig.~3):
 
\begin{equation} 
S_2(\eta) = \frac{S_1(\eta)}{r_{\rm t}} \propto 
a^{-1}\eta^{-2} \sqrt{1 - e^2} \left(2\eta - (1 + \eta^2) \sqrt{1-e^2}\right)^{-\frac{1}{2}} ~,  
\end{equation} 
 
\noindent 
where we have used Eq.~15 and Eq.~6 to describe $r_{\rm t}$ in terms of
$\eta$.  The average $\eta$ that can be expected is then:
 
\begin{equation} 
<\eta> = \frac{\int_{n_{\rm min}}^{n_{\rm max}}\eta S_2(\eta) 
d\eta}{\int_{n_{\rm min}}^{n_{\rm max}}S_2(\eta) d\eta} = \sqrt{1-e^2} 
\end{equation} 
 
\noindent 
where $n_{\rm max} = \frac{\sqrt{1-e^2}}{1-e}$ and $n_{\rm min} =
\frac{\sqrt{1-e^2}}{1+e}$, as defined by Eq.~15. Thus, we can expect the
average observed $\eta$ to decrease with increasing
eccentricity. Unfortunately, the same is also true for binary stars, which
increases the chance of a non-planetary system producing a planet-like
diagnostic.
 
\section{Putting the Diagnostic to the Test}

We now illustrate the effectiveness of the diagnostic $\eta$ in separating
exoplanetary-like transit candidates from a large fraction of the pretenders
that may be caused by eclipsing binary stars, which are either grazing or
heavily blended.

\subsection{Actual Transit Candidates} 
 
The transit candidates discovered by the OGLE group
\citep{udalski1,udalski2,udalski3,udalski4,udalski5} provide an excellent
opportunity to test the performance of the exoplanet diagnostic.  Orbital
periods, transit durations and depths for 137 candidates are available, along
with a short list of confirmed exoplanets in the sample. The results of the
application of the exoplanet diagnostic ($\eta_{\rm p}$) to the OGLE
candidates are shown in Fig.~4 (and the data used to create it are found in
Table 1).We have assumed a planetary radius of $1 R_{\rm J}$, as
representative of the largest hot Jupiters found. For comparison, modeled
central transits of Jupiter and Saturn and the actual transit data of HD209458
\citep{brown} and trES-1 \citep{alonso} are included.

Of those OGLE candidates that have been followed up with further observations
(OGLE-TR-59 and lower), OGLE-TR-56 \citep{konacki1} and OGLE-TR-10
\citep{bouchy} are the only two confirmed exoplanets in this sample. Indeed,
OGLE-TR-10 was confirmed after we identified it as interesting from our
diagnostic analysis. They have the lowest and fifth lowest exoplanet
diagnostics, respectively. The exoplanets that have been identified in the
later sample (OGLE-TR-111 \citep{pont}, OGLE-TR-113 and OGLE-TR-132
\citep{bouchy}), which has not been studied completely (OGLE-TR-60 and
higher), are also among those with the lowest diagnostics. The modeled Jupiter
and Saturn also produce low diagnostics, as do HD209458 \citep{brown} and
TrES-1 \citep{alonso}. This evidence makes a strong argument that our
diagnostic can effectively identify candidates that are more likely to be
exoplanets. As such, it seems likely that other exoplanets will be discovered
in this second set of OGLE candidates. Additionally, our analysis suggests
that several OGLE candidates warrant (see Table 1) a more thorough
investigation, while other disputed events, such as OGLE-TR-33, do not. We
note, in fact, that the most recent research suggests that OGLE-TR-33 is in
fact a blend \citep{torres}.

\subsection{Modeled Blended Stellar Binaries}
 
Testing the response of the exoplanet diagnostic to blended eclipsing binary
systems is not as simple.  In order to get an accurate impression of the
events that could be observed to mimic exoplanetary transits, many different
types of binary systems must be modeled at a variety of projected
inclinations. The types of exoplanets that could be observed must also be
modeled. Both of these tasks can be readily performed using the expectations
from a standard zero-age main sequence (ZAMS) and limb darkening laws -- in
this case, we used those from \citet{claret}. The blends can then be
constructed using the modeled binary star transits and the equations for
blends described by \citet{tingley}. Using $\eta_\star$, the diagnostics of
modeled blends can be compared to those of modeled exoplanets.
 
Other methods may be used to eliminate blends. For example, blends will have
V-shaped transits or secondary transits that would be readily identifiable in
the light curves. As with grazing binaries, blends can also have primary and
secondary transits of almost the same depth, again leading to an erroneous
period and an increased exoplanet diagnostic. In what follows, we ignore these
effects, concentrating on information from the diagnostic alone, and assuming
circular orbits.

Figures 5-7 show how the diagnostic $\eta_\star$ evolves for blends compared
to exoplanets for a variety of assumptions and conditions. The modeled blends
are created by combining modeled light curves of blends and adding the proper
amount of light from the `primary'' (the constant, brighter component for
blends and the parent star for exoplanets) to create a transit of the desired
depth. These figures are designed to illustrate the overall effectiveness of
the diagnostic under different conditions. When studying these figures, it is
important to remember that the spectral type of the primary component of a
blend has nothing to do with the spectral type of the eclipsing component --
it only provides photons to make the observed transits shallower.

Figure 5 shows how the diagnostic $\eta_\star$ evolves for blends relative to
exoplanets for different spectral types and transit percentages, assuming that
nothing is known about the ``primary'' and therefore using $R_\star = R_\odot$
when calculating $\eta_\star$. The lines help to illustrate the likelihood of
finding a blends with the listed spectral type and transit percentage that
have similar diagnostic to exoplanets. The shaded regions show what
diagnostics exoplanets with radii from $1.5 R_{\rm J}$ (right boundary) to
$R_{\rm Mercury}$ (left boundary) and transit percentages from 0\% (top) to
75\% (bottom) would produce if the ``primary'' was an F2 star (angled $45^o$
to the right), a Solar-type star (horizontal dotted lines) and an M3 star
(angled $45^o$ to the left). Unblended binaries occupy the right-hand end of
each line. From this figure, it is clear that in most cases, even without any
knowledge of the ``primary'', most blends are distinguishable from exoplanets
in the case of circular orbits, even for exoplanets smaller than Earth. This
is especially true for blends involving earlier-type eclipsing binaries, as
those tend to have relatively long transits.
 
Figure 6 shows how $\eta_\star$ is influenced specifically by the spectral
type of the binary in the blend for different transit percentages. Here too it
is evident that later main-sequence spectral types in the binary produce lower
diagnostics. As this is due to the shorter transit durations (from the lower
stellar radii), giant and sub-giant binaries in blends would produce very high
diagnostics that would be easily separable from exoplanets. The shaded
regions in this figure are identical to those in Fig.~5. This figure also
shows that a relatively high percentage of blends can be separated from
exoplanets, especially those orbiting late spectral-type parent stars.
 
Incorporating an estimate of the primary radius improves the performance of
the diagnostic. Figure 7 is similar to Figure 6, except that stellar radii are
assumed to be known to within 20\% (top) and 5\% (bottom).  For clarity, only
the highest (F2/F2 binary) and the lowest (M3/M3 binary) of the 70\% transit
percentage contours are shown. Information on stellar radii affects the
locations of the exoplanet regions so that central transits have a diagnostic
close to one, within the limits of the errors, which is quite different from
earlier figures. Moreover, the imprecision in the estimate of stellar radius
affects the diagnostics of the blends as well.  In this figure, the
information on spectral type show that blends in which the ``primary'' is a
late spectral type will be the easiest to separate from exoplanets, while
those involving the earliest spectral types will not gain much from the use of
the diagnostic. To date, only about 12\% of exoplanets discovered with either
radial velocity or transits have been around F stars. This of course could be
affected by the choice of stars in radial velocity searches.

\section{Conclusions} 
 
We have devised an exoplanet diagnostic $\eta$ and shown that it can be an
effective tool in choosing transit candidates worthy of follow-up observation.
Using our diagnotic, many eclipsing binary stars, especially those involving
giants, and blends can be excluded from further consideration. Specifically,
we suggest that OGLE-TR-33 is unlikely to be planetary, while our analysis
indicated that OGLE-TR-10 had a high likelihood of begin planetary before it
was confirmed \citep{bouchy}. Estimates of stellar radii improve the
performance of the diagnostic, giving it an even greater ability to identify
the events that are non-exoplanetary in origin.  The diagnostic will be
particularly useful for bodies (such as hot Jupiters) that have circular
orbits.  The addition of eccentricity to the analysis leads to an increased
spread in the diagnostic, but does not critically impair the method.  Transits
are more likely to occur in portions of an eccentric orbit where $v_{\rm t}$
exceeds $v_{\rm circular}$, leading to a reduction in the average diagnostic
for exoplanetary systems with a given period. Overall, we encourage the use of
this diagnostic tool in the search for exoplanets via transits to reduce the
large numbers of candidates that require resource-consuming follow-up
observations.
 
%% \acknowledgments 

\clearpage 
 
\begin{figure} 
  \plottwo{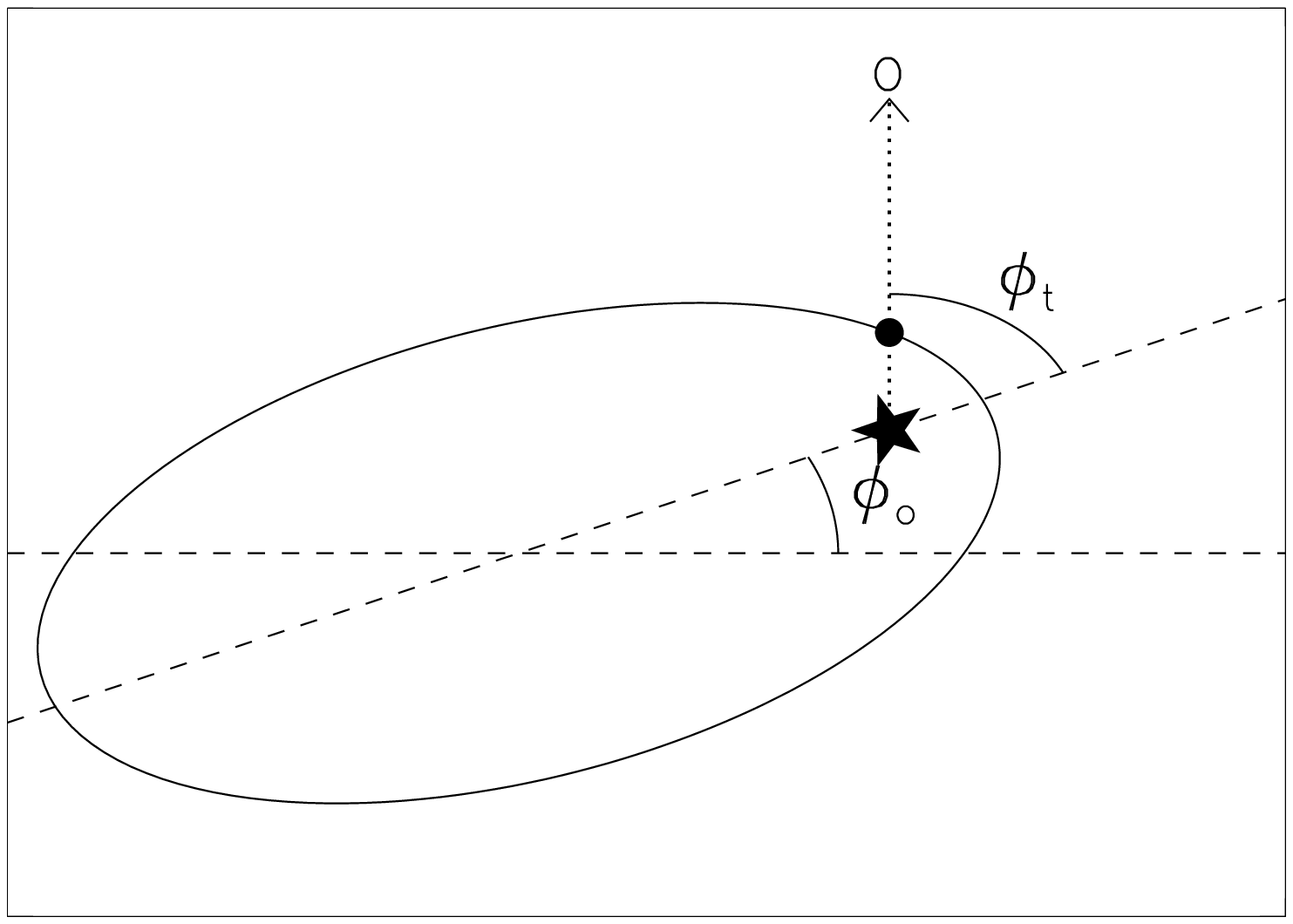}{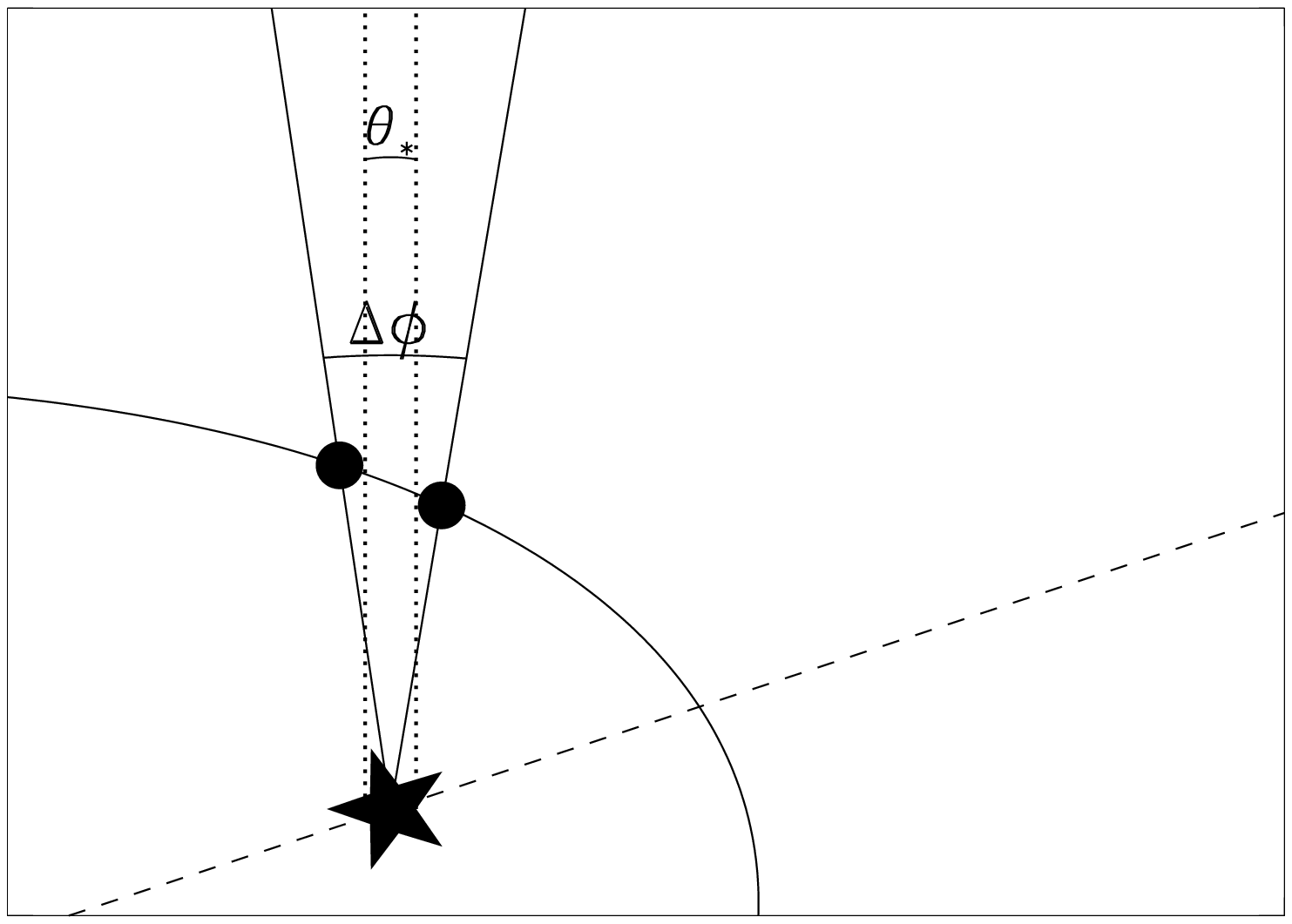} 
  \caption{Geometry of an observed transit and definitions of relevant angles
  used in the text. $\frac{\pi}{2} - \phi_{\rm o}$ is the angle between the
  semi-major axis of the orbit and the line of sight. $\phi_{\rm t}$ is the
  eccentric angle of transit center relative to $\phi_{\rm o}$. $\theta_*$ is
  the angular size of the parent star as measured by a distant observer. The
  eccentric angle $\Delta\phi$ is measured between first contact and last
  contact of the transit.}
\end{figure} 

\begin{figure}
   \plotone{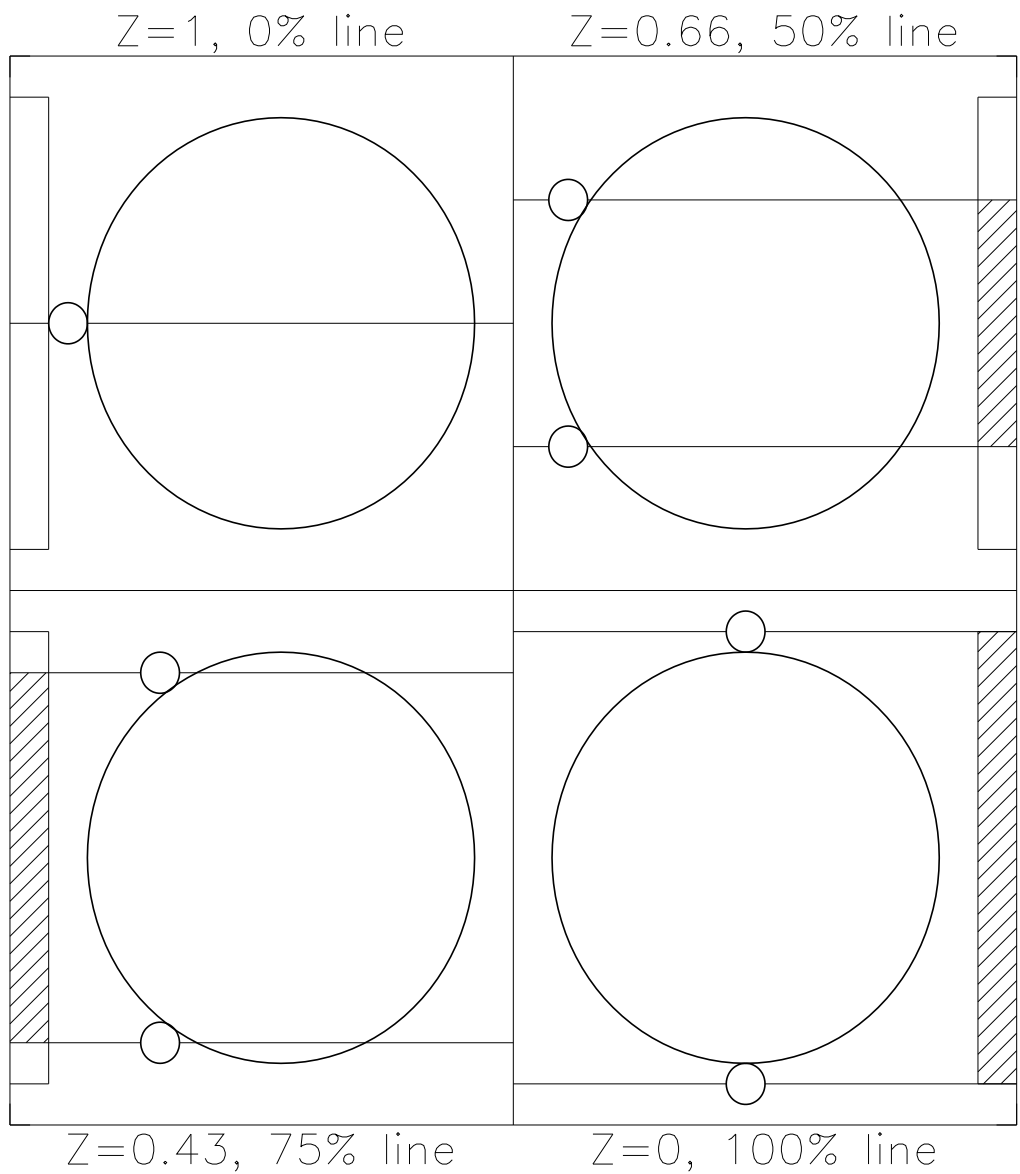}
   \caption{A visualization of the geometric interpretation
   of $Z$. This figure demonstrates how $Z$ relates to the path of an
   exoplanets across the disk and how that, in turn, relates to the percentage
   of all possible transits (assuming random orbital orientations) that will
   exhibit a $Z$ greater than or equal to a chosen $Z$. The large circles
   represent the parent stars while the small ones are exoplanets. The
   straight lines across the disks are the orbital paths of the exoplanets,
   which correspond to the $Z$s shown at the top of the figure. The boxes at
   the outside edge of the figure show what percentage of all possible
   transits (assuming random orbits) that have a $Z$ greater than the listed
   $Z$. This percentage (which will be referred to later as transit percentage
   ($P_t$)) is related to $Z$ by $P_t=sqrt{1-Z^2}$. The upper-left
   hand figure shows a central transit (with $Z=1$ and the bottom-right shows
   the other extreme, the narrowest of grazing transits, corresponding to
   $Z=0$. The other two figures show in-between cases, demonstrating how
   transit duration (proportional to the crossing time across the disk of the
   star from first contact to last) varies with $Z$.}
\end{figure}
    
\begin{figure} 
   \plotone{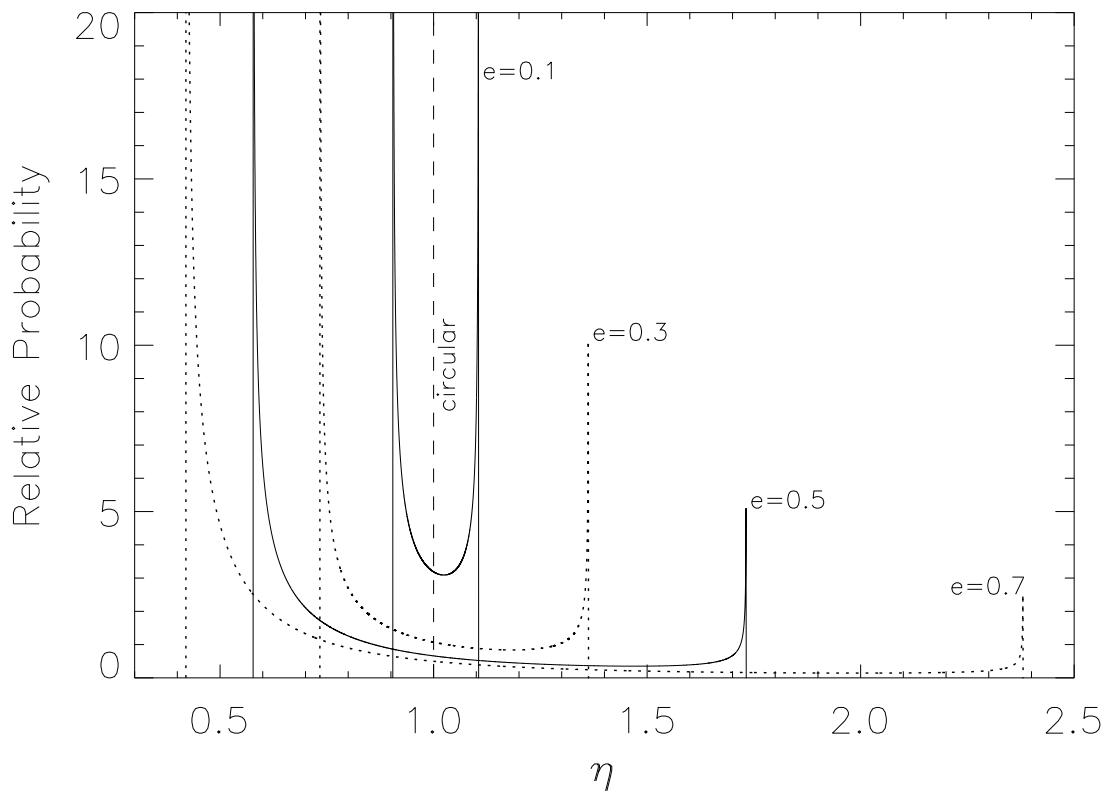} 
   \caption{The effect of eccentricity on the spread of possible diagnostics
   $\eta$ for a central transit with a fixed period. The labels on each line
   represent the eccentricity and the dashed line represents the exoplanet
   diagnostic for a circular orbit. Notice how the spread increases as the
   eccentricity increases, but the probability of a transit having a large
   diagnostic decreases significantly ($\propto \sqrt{1-e^2}$ as demonstrated
   in Eq.~18). As the diagnostics shown are only for central transits, they
   therefore represent the maximum possible observed diagnostic for the given
   parameters.}
\end{figure} 
 
\begin{figure} 
   \plotone{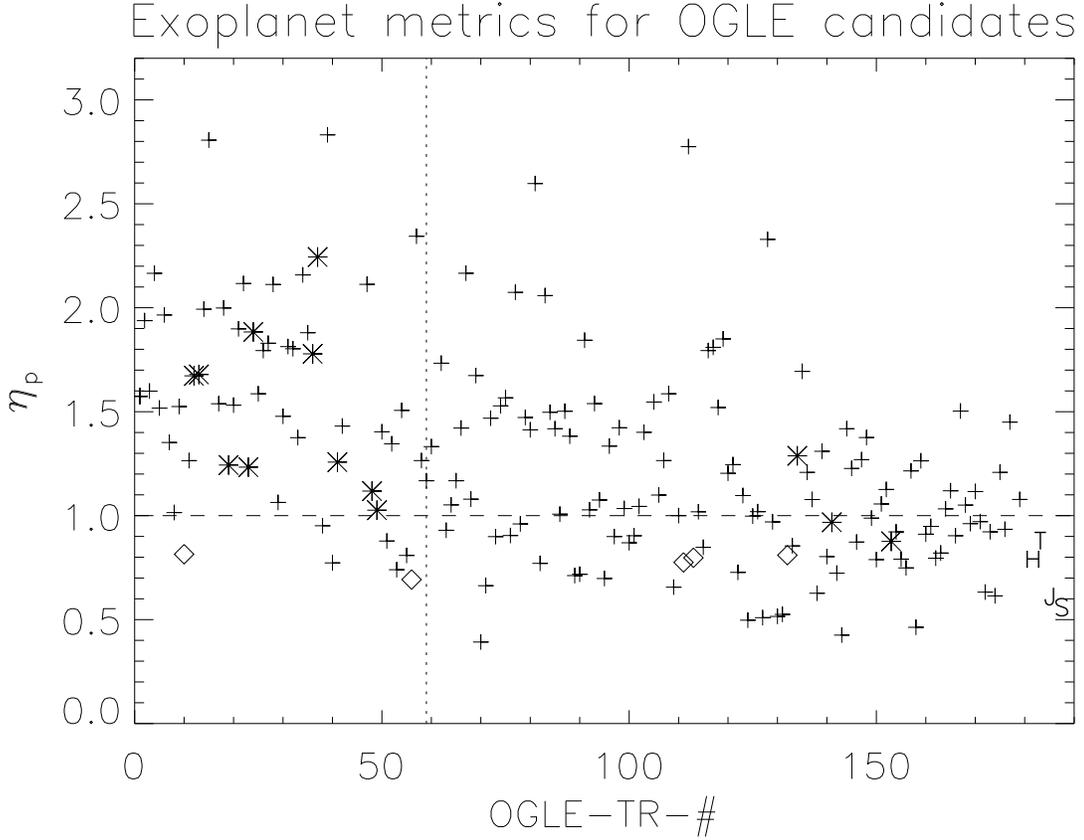} 
   \caption{The exoplanet diagnostic plotted against OGLE transit candidate
   identifiers, assuming an exoplanetary radius of $1 R_J$. The horizontal
   dashed line represents the cutoff below which any exoplanets in the sample
   should fall. The vertical dotted line shows the dividing line between the
   subset that is completely followed-up (left side) and those for which the
   analysis is incomplete (right side). The crosses represent candidates that
   are either binary stars or unclassified, while the diamonds are the known
   exoplanets in the sample.  The observed transit of HD209458 \citep{brown}
   is marked with an ``H'' and the observed transit of TrES-1 \cite{alonso}
   marked with a ``T''.  A theoretical central Jupiter transit is marked with
   a ``J'' and a theoretical central Saturn transit is marked with an
   ``S''. Note that the exoplanets (both known and modeled) lie well below the
   cut-off line given by our criterion $\eta \leq 1$. Lastly, the observed
   slope in this figure is not real; it is likely indicative of the OGLE group's
   increasing ability to remove eclipsing binaries from the candidate
   list. The data used to create this figure can be found in Table 1.} 
\end{figure} 
 
\begin{figure} 
   \plotone{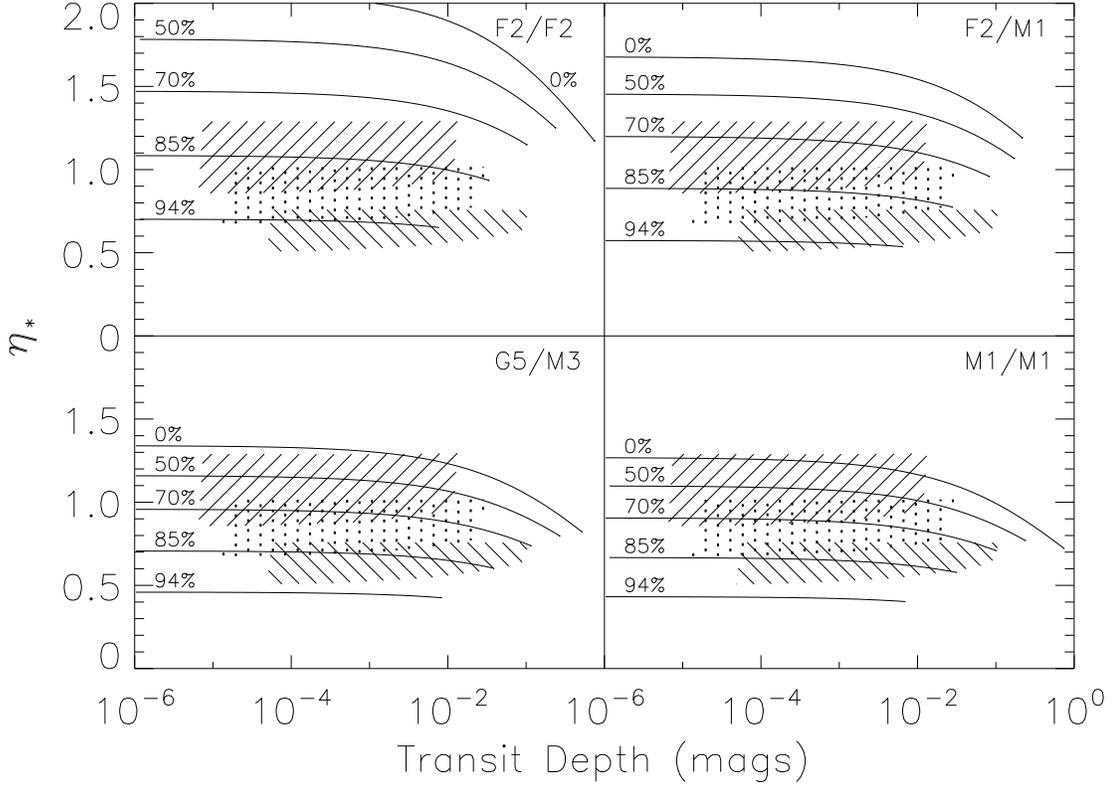}
   \caption{The exoplanet diagnostic for different blended
   binary systems at transit percentages (as described in Fig.~2) as a
   function of observed transit depth, assuming that the stellar radius for
   the ``primary'' (the constant component for a blend, the parent star for an
   exoplanet) is unknown and therefore using $R_\odot$ to calculate
   $\eta_\star$. The lines depict the evolution of the
   diagnostic of a series of blends as increasingly more light from a
   unvarying third component is added to the light from an eclipsing binary to
   reduce the transit depth.
   %The numbers show the
   %percentage of blends where the eclipsing component is comprised of the
   %listed spectral types lie above the associated line. The lines are created
   %by modeling the binary system and gradually adding more and more light from
   %a non-eclipsing third component, which reduces the transit depth. 
   The lines are labeled by the transit percentage. For comparison,
   shaded regions indicate diagnostics for exoplanets ranging in size from
   $1.5 R_J$ down to $R_{\rm Mercury}$ around F2 stars (highest), the Sun
   (middle) and an M3 star (bottom), assuming that $R_\star$ is known when
   calculating $\eta_\star$. A shaded region is bounded on top by modeled
   central transits and on the bottom by modeled transits with a transit
   percentage of 75\%, or $Z=sqrt{1.-0.75^2}=0.66$. Note that the
   majority of blends are not in regions where exoplanets would be found for
   larger stars. True grazing binaries without blending fall where lines end
   on the right.}
\end{figure} 
 
\begin{figure} 
   \plotone{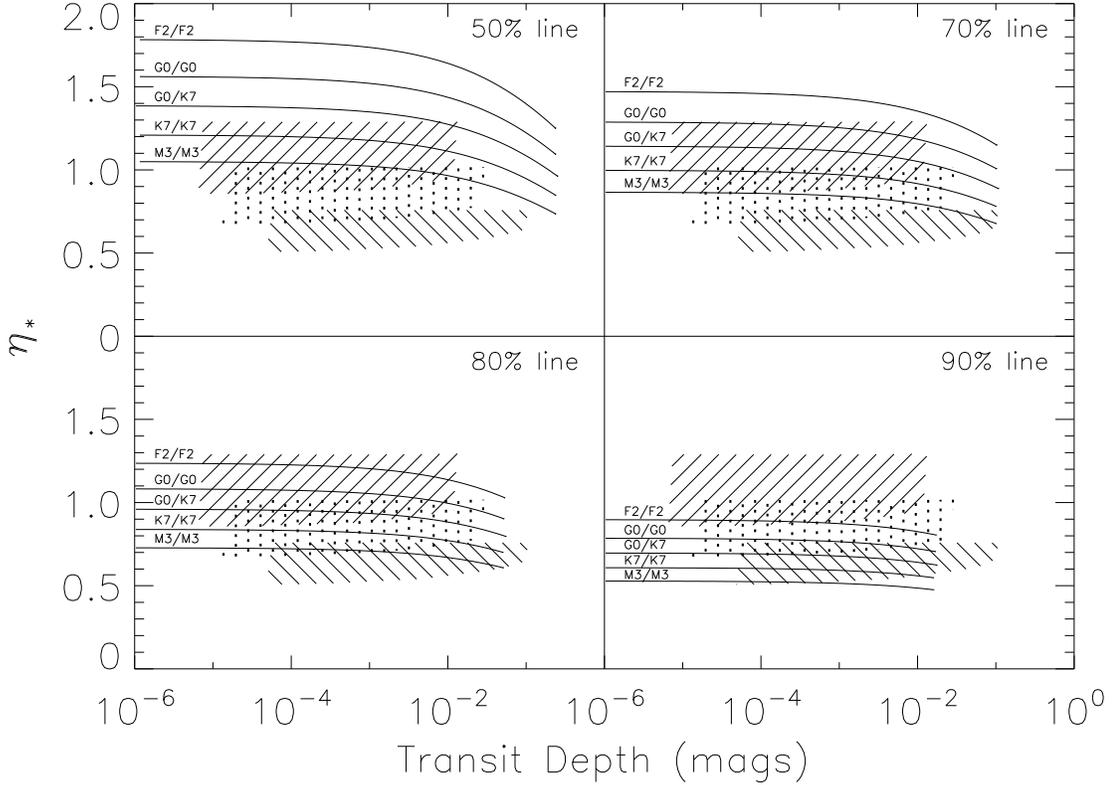} 
   \caption{Similar to Fig.~5, this figure shows how the diagnostics of
   blends for different transit percentages and binary spectral types compare
   to those for exoplanets, again assuming $R_\star = R_\odot$.
   The label for each line indicates the spectral
   types modeled, while the label for each subsection of the figure
   corresponds to the inclination. The shaded regions are as described in
   Fig.~5. This figure demonstrates that the diagnostic is effective for
   separating most, but certainly not all, blends from exoplanets. The
   confusion will be most acute for those exoplanets around the more massive
   stars included in the study.}
%   {Contours representing the 50\%, 70\%, 80\% and 90\% occurrence
%   probabilities (i.e. 50\%-90\% of blends, respectively, will fall above that
%   line) for blends from a large variety of different possible binary systems,
%   ranging from an F2/F2 system, which produced the highest diagnostics of all
%   the systems modeled, to M3/M3, which produced the lowest. The shaded
%   regions are as described in Fig.~4.  Notice that about half of the shaded
%   region for exoplanets orbiting solar-type stars lies below all the 70\%
%   lines. However, it is also clear that not all blends can be ruled out using
%   the exoplanet diagnostic.}
\end{figure} 
 
\begin{figure} 
   \plotone{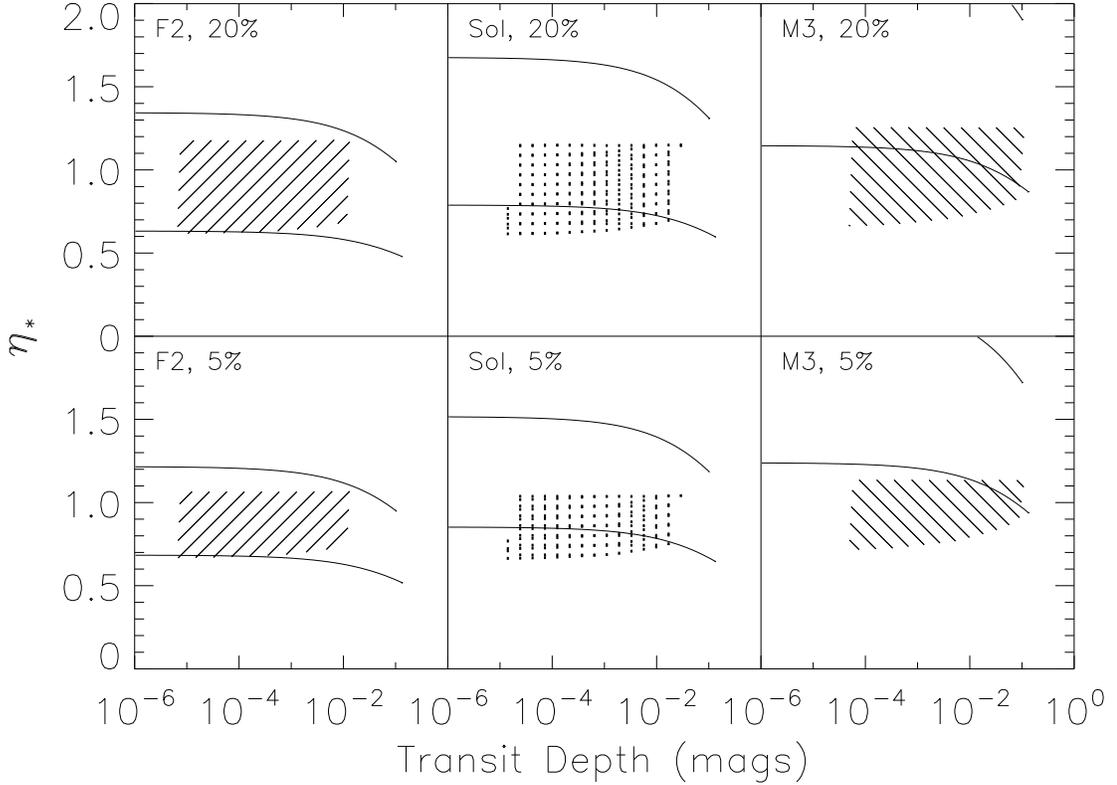}
   \caption{Similar to Fig.~6, except for the fact that
   stellar radius for the ``primary'' is known to either 20\% (top) or 5\%
   (bottom) precision when calculating $\eta_\star$, the ``primary'' being the
   constant component for blends and the parent star for exoplanets. Only 70\%
   contours for a blend with a F2/F2 (highest modeled diagnostics) and an
   M3/M3 (lowest modeled diagnostics) eclipsing component are shown for
   clarity. The shades regions show the range of diagnostics for exoplanets
   (75\% transit percentages) around the stars of the listed spectral types
   where the radius is known to the listed precision. Notice how the shades
   regions cover approximately the same range in diagnostics in all the
   figures, though narrower for the bottom three figures due to the higher
   precision in the measured ``primary'' radius.  Notice also how exoplanets
   are easier to separate from blends for later spectral types.}
\end{figure} 

\begin{deluxetable}{lcccc}
%\tabletypesize{8pt}
%\rotate
\tablewidth{0pt}
%\tablenum{}
  \tablecolumns{5}
  \tablecaption{Transit Parameters and Diagnostics for the OGLE candidates}
  \tablehead{\colhead{OGLE-TR-\#} & \colhead{Period (days)} &
  \colhead{Duration (days)} & \colhead{Depth(mags)} & \colhead{$\eta_{\rm p}$} } 
  \startdata
1    &         1.600  &     0.117  &   0.043     &        1.02  \\  
2    &         2.813  &     0.209  &   0.019     &        1.54  \\  
3    &         1.189  &     0.129  &   0.019     &        1.27  \\  
4    &         2.618  &     0.174  &   0.065     &        2.06  \\  
5    &         0.808  &     0.090  &   0.043     &        1.08  \\  
6    &         4.548  &     0.197  &   0.053     &        1.78  \\  
7    &         2.717  &     0.126  &   0.034     &        1.38  \\  
8    &         2.715  &     0.088  &   0.048     &        0.99  \\  
9    &         3.268  &     0.140  &   0.048     &        1.61  \\  
10    &        3.101  &     0.091  &   0.019     &       0.82  \\ 
11    &        1.615  &     0.090  &   0.053     &       1.07  \\ 
12    &        5.772  &     0.195  &   0.038     &       1.01  \\ 
13    &        5.853  &     0.208  &   0.030     &       1.49  \\ 
14    &        7.797  &     0.264  &   0.034     &       2.09  \\ 
15    &        4.874  &     0.337  &   0.026     &       2.56  \\ 
16    &        2.138  &     0.451  &   0.026     &       2.21  \\ 
17    &        2.317  &     0.136  &   0.034     &       1.60  \\ 
18    &        2.228  &     0.165  &   0.043     &       1.27  \\ 
19    &        5.282  &     0.126  &   0.065     &       1.02  \\ 
20    &        4.283  &     0.147  &   0.059     &       1.31  \\ 
21    &        6.892  &     0.229  &   0.043     &       2.70  \\ 
22    &        4.275  &     0.180  &   0.084     &       2.35  \\ 
23    &        3.286  &     0.109  &   0.059     &       1.17  \\ 
24    &        5.282  &     0.199  &   0.053     &       1.77  \\ 
25    &        2.218  &     0.135  &   0.038     &       1.44  \\ 
26    &        2.538  &     0.148  &   0.053     &       1.39  \\ 
27    &        1.714  &     0.155  &   0.026     &       1.75  \\ 
28    &        3.405  &     0.193  &   0.053     &       2.06  \\ 
29    &        2.715  &     0.092  &   0.048     &       0.99  \\ 
30    &        2.365  &     0.128  &   0.038     &       1.50  \\ 
31    &        1.883  &     0.154  &   0.030     &       1.71  \\ 
32    &        1.343  &     0.133  &   0.034     &       0.93  \\ 
33    &        1.953  &     0.115  &   0.034     &       1.34  \\ 
34    &        8.581  &     0.273  &   0.048     &       2.13  \\ 
35    &        1.259  &     0.139  &   0.030     &       1.10  \\ 
36    &        6.251  &     0.194  &   0.059     &       1.80  \\ 
37    &        5.719  &     0.275  &   0.030     &       2.20  \\ 
38    &        4.101  &     0.094  &   0.048     &       1.14  \\ 
39    &        0.815  &     0.181  &   0.030     &       1.00  \\ 
40    &        3.430  &     0.083  &   0.026     &       0.76  \\ 
41    &        4.517  &     0.153  &   0.022     &       1.46  \\ 
42    &        4.161  &     0.149  &   0.038     &       1.80  \\ 
47    &        2.335  &     0.214  &   0.019     &       1.73  \\ 
48    &        7.225  &     0.159  &   0.022     &       1.00  \\ 
49    &        2.690  &     0.095  &   0.034     &       1.03  \\ 
50    &        2.248  &     0.114  &   0.048     &       1.25  \\ 
51    &        1.748  &     0.071  &   0.034     &       0.85  \\ 
52    &        1.325  &     0.099  &   0.034     &       1.28  \\ 
53    &        2.905  &     0.070  &   0.034     &       0.84  \\ 
54    &        8.162  &     0.215  &   0.026     &       1.35  \\ 
55    &        3.184  &     0.091  &   0.019     &       1.00  \\ 
56    &        1.211  &     0.062  &   0.013     &       0.64  \\ 
57    &        1.674  &     0.197  &   0.026     &       1.72  \\ 
58    &        4.345  &     0.142  &   0.030     &       1.76  \\ 
59    &        1.497  &     0.095  &   0.026     &       1.02  \\ 
60    &        2.308  &     0.140  &   0.016     &       1.21  \\ 
61    &        4.268  &     0.474  &   0.030     &       3.33  \\ 
62    &        2.601  &     0.155  &   0.038     &       1.83  \\ 
63    &        1.066  &     0.083  &   0.011     &       0.85  \\ 
64    &        2.717  &     0.108  &   0.022     &       0.93  \\ 
65    &        0.860  &     0.074  &   0.034     &       1.01  \\ 
66    &        3.514  &     0.131  &   0.053     &       1.27  \\ 
67    &        5.279  &     0.229  &   0.053     &       1.97  \\ 
68    &        1.288  &     0.081  &   0.030     &       1.02  \\ 
69    &        2.337  &     0.145  &   0.038     &       1.28  \\ 
70    &        8.040  &     0.048  &   0.053     &       0.36  \\ 
71    &        4.187  &     0.079  &   0.022     &       0.69  \\ 
72    &        6.854  &     0.173  &   0.048     &       1.38  \\ 
73    &        1.581  &     0.070  &   0.034     &       0.88  \\ 
74    &        1.585  &     0.122  &   0.030     &       0.78  \\ 
75    &        2.642  &     0.144  &   0.034     &       1.58  \\ 
76    &        2.126  &     0.086  &   0.022     &       0.70  \\ 
77    &        5.455  &     0.269  &   0.022     &       1.90  \\ 
78    &        5.320  &     0.115  &   0.030     &       1.05  \\ 
79    &        1.324  &     0.111  &   0.030     &       1.18  \\ 
80    &        1.807  &     0.137  &   0.016     &       1.25  \\ 
81    &        3.216  &     0.282  &   0.022     &       1.86  \\ 
82    &        0.764  &     0.047  &   0.034     &       0.76  \\ 
83    &        1.599  &     0.191  &   0.016     &       0.78  \\ 
84    &        3.113  &     0.129  &   0.059     &       1.41  \\ 
85    &        2.114  &     0.113  &   0.048     &       1.57  \\ 
86    &        2.777  &     0.082  &   0.065     &       0.93  \\ 
87    &        6.606  &     0.167  &   0.059     &       1.52  \\ 
88    &        1.250  &     0.099  &   0.034     &       1.37  \\ 
89    &        2.289  &     0.078  &   0.013     &       0.73  \\ 
90    &        1.041  &     0.054  &   0.022     &       0.55  \\ 
91    &        1.579  &     0.136  &   0.043     &       1.52  \\ 
92    &        0.978  &     0.066  &   0.038     &       0.95  \\ 
93    &        2.206  &     0.153  &   0.019     &       1.40  \\ 
94    &        3.092  &     0.099  &   0.043     &       1.05  \\ 
95    &        1.393  &     0.059  &   0.019     &       1.16  \\ 
96    &        3.208  &     0.125  &   0.043     &       1.08  \\ 
97    &        0.567  &     0.059  &   0.016     &       0.95  \\ 
98    &        6.398  &     0.176  &   0.034     &       1.52  \\ 
99    &        1.102  &     0.071  &   0.034     &       0.88  \\ 
100    &       0.826  &     0.062  &   0.019     &      0.82  \\
101    &       2.361  &     0.078  &   0.038     &      0.75  \\
102    &       3.097  &     0.116  &   0.019     &      0.88  \\
103    &       8.216  &     0.175  &   0.048     &      1.55  \\
104    &       6.068  &     0.413  &   0.053     &      1.94  \\
105    &       3.058  &     0.159  &   0.026     &      1.05  \\
106    &       2.535  &     0.110  &   0.022     &      1.09  \\
107    &       3.189  &     0.113  &   0.053     &      1.26  \\
108    &       4.185  &     0.158  &   0.048     &      1.49  \\
109    &       0.589  &     0.052  &   0.008     &      0.85  \\
110    &       2.848  &     0.100  &   0.026     &      0.90  \\
111    &       4.016  &     0.094  &   0.019     &      0.78  \\
112    &       3.879  &     0.346  &   0.016     &      2.45  \\
113    &       1.432  &     0.061  &   0.030     &      0.80  \\
114    &       1.712  &     0.086  &   0.026     &      0.88  \\
115    &       8.346  &     0.102  &   0.059     &      0.38  \\
116    &       6.064  &     0.184  &   0.077     &      1.87  \\
117    &       5.022  &     0.213  &   0.030     &      1.52  \\
118    &       1.861  &     0.143  &   0.019     &      1.65  \\
119    &       5.282  &     0.210  &   0.038     &      1.84  \\
120    &       9.165  &     0.142  &   0.077     &      1.23  \\
121    &       3.232  &     0.105  &   0.071     &      1.28  \\
122    &       7.268  &     0.107  &   0.019     &      0.78  \\
123    &       1.803  &     0.126  &   0.008     &      1.12  \\
124    &       2.753  &     0.061  &   0.011     &      0.53  \\
125    &       5.303  &     0.145  &   0.013     &      1.01  \\
126    &       5.110  &     0.129  &   0.022     &      0.95  \\
127    &       1.927  &     0.055  &   0.011     &      0.60  \\
128    &       7.391  &     0.360  &   0.016     &      2.26  \\
129    &       5.740  &     0.116  &   0.034     &      0.76  \\
130    &       4.830  &     0.061  &   0.028     &      0.49  \\
131    &       1.869  &     0.056  &   0.011     &      0.58  \\
132    &       1.689  &     0.084  &   0.011     &      0.80  \\
133    &       5.310  &     0.099  &   0.034     &      0.81  \\
134    &       4.537  &     0.186  &   0.011     &      0.87  \\
135    &       2.573  &     0.177  &   0.019     &      1.56  \\
136    &       3.115  &     0.140  &   0.016     &      1.22  \\
137    &       2.537  &     0.101  &   0.030     &      0.96  \\
138    &       2.645  &     0.069  &   0.016     &      0.86  \\
139    &       2.534  &     0.122  &   0.030     &      0.96  \\
140    &       3.393  &     0.076  &   0.043     &      0.79  \\
141    &       5.678  &     0.132  &   0.019     &      0.74  \\
142    &       3.062  &     0.070  &   0.034     &      0.56  \\
143    &       3.349  &     0.065  &   0.006     &      0.39  \\
144    &       2.445  &     0.127  &   0.034     &      1.24  \\
145    &       2.741  &     0.131  &   0.019     &      1.21  \\
146    &       2.944  &     0.079  &   0.043     &      0.95  \\
147    &       3.841  &     0.136  &   0.030     &      0.99  \\
148    &       1.432  &     0.114  &   0.022     &      1.25  \\
149    &       4.551  &     0.116  &   0.026     &      0.86  \\
150    &       2.073  &     0.107  &   0.005     &      0.93  \\
151    &       1.483  &     0.089  &   0.022     &      1.01  \\
152    &       3.730  &     0.133  &   0.019     &      1.12  \\
153    &       4.395  &     0.098  &   0.030     &      0.85  \\
154    &       3.669  &     0.119  &   0.013     &      0.90  \\
155    &       5.277  &     0.130  &   0.008     &      1.01  \\
156    &       3.583  &     0.076  &   0.034     &      0.64  \\
157    &       5.868  &     0.139  &   0.043     &      0.91  \\
158    &       6.384  &     0.065  &   0.019     &      0.39  \\
159    &       2.126  &     0.106  &   0.038     &      1.26  \\
160    &       4.901  &     0.146  &   0.008     &      0.72  \\
161    &       2.747  &     0.135  &   0.006     &      1.03  \\
162    &       3.758  &     0.107  &   0.011     &      0.66  \\
163    &       0.946  &     0.053  &   0.034     &      0.80  \\
164    &       2.681  &     0.109  &   0.019     &      0.95  \\
165    &       2.891  &     0.101  &   0.043     &      1.33  \\
166    &       5.219  &     0.136  &   0.011     &      0.88  \\
167    &       5.261  &     0.192  &   0.022     &      0.65  \\
168    &       3.650  &     0.135  &   0.013     &      0.67  \\
169    &       2.768  &     0.103  &   0.019     &      1.28  \\ 
170    &       4.136  &     0.127  &   0.026     &      1.22  \\ 
171    &       2.091  &     0.081  &   0.038     &      0.99  \\
172    &       1.793  &     0.078  &   0.006     &      0.74  \\
173    &       2.605  &     0.097  &   0.019     &      0.49  \\
174    &       3.110  &     0.071  &   0.016     &      0.74  \\
175    &       1.488  &     0.105  &   0.019     &	1.15  \\
176    &       3.404  &     0.103  &   0.022     &      0.87  \\
177    &       5.643  &     0.168  &   0.038     &      0.81  \\
HD209458   &   3.524  &     0.1298 &   0.0167	 &      0.85  \\
TrES-1   &     3.030  &     0.1250 &   0.023     &	 0.94  \\
Jupiter   &     4346  &     1.236  &   0.010     &	 0.67  \\
Saturn   &     10760  &     1.675  &   0.0073    &	 0.62  \\
    \enddata

\tablecomments{Note that the durations were measured estimated using
a simple matched filter code from the data available on the OGLE
website, http://bulge.princeton.edu/~ogle/.}
\end{deluxetable}

\end{document}